\documentclass{INTERSPEECH2023}

\interspeechcameraready

\title{Video Multimodal Emotion Recognition System for Real World Applications}
\name{Sun-Kyung Lee and Jong-Hwan Kim$^{*}$\thanks{*Corresponding Author}}
\address{KAIST, Daejeon, Republic of Korea}
\email{\{sklee, johkim\}@rit.kaist.ac.kr}

\begin{document}

\maketitle
 
\begin{abstract}
This paper proposes a system capable of recognizing a speaker\textquotesingle{}s utterance-level emotion through multimodal cues in a video. The system seamlessly integrates multiple AI models to first extract and pre-process multimodal information from the raw video input. Next, an end-to-end MER model sequentially predicts the speaker\textquotesingle{}s emotions at the utterance level. Additionally, users can interactively demonstrate the system through the implemented interface.
\end{abstract}

\section{Introduction}
\label{sec:intro}

Emotions have a significant impact on people in their daily lives. Individuals perceive and understand the emotions of others through three major modalities; facial expressions (visual), paralinguistics (acoustic) and spoken words (textual), This information is crucial for building social relationships. As videos are gaining more and more popularity as a means of communication, precisely capturing the emotions of targets using these multimodal cues is crucial for active human-computer interaction applications. With recent development of deep neural networks, various deep learning based multimodal emotion recognition (MER) models have been presented. Most of the existing works focused on research-level fusion methods of the different multimodal inputs. However, there have been only a few attempts to develop actually operational systems for real world applications \cite{wei2022fv2es}. 

To tackle this issue, we propose an MER system capable of capturing a speaker\textquotesingle{}s emotion from a raw video of any length. Since processing the entire video at once is computationally infeasible, the system initially performs utterance-level pre-processing. Specifically, the video is segmented into multiple clips containing each utterance, and multimodal information for MER is extracted respectively. Subsequently, an end-to-end MER model sequentially takes the pre-processed unaligned visual, acoustic and textual modalities, and predicts the speaker\textquotesingle{}s emotion at the utterance level. This approach enables users to trace changes in the speaker\textquotesingle{}s emotions at each utterance. Additionally, by averaging all the utterance-level predictions, the overall emotions of the entire video can be observed. We also provide an interface that enables users to demonstrate the system by uploading videos.

\section{End-to-End MER Model}

\subsection{Model Design}

Previously, the majority of approaches to MER were two-stage, involving the extraction of fixed features from each modality using hand-crafted algorithms in the first stage, followed by their fusion in the second stage. One limitation of this method was that the extracted features were not fine-tuned for the task of MER. Consequently, recent works have made a progress by employing an end-to-end learning approach for joint optimization of feature extraction and fusion.

One important thing to note is that gathering a sufficient amount of MER data to train a large model can be costly. To tackle this issue, we leveraged modal-specific large pre-trained models for each modality and jointly trained them with a feature fusion component. With recent success of large pre-trained models and transfer learning, the integrated model can fully leverage modal-specific knowledge, even with a relatively small amount of  MER data.

To be more specific, InceptionResNet\footnote{\label{note}https://github.com/timesler/facenet-pytorch} pre-trained on VGGFace2, DistilHuBERT \cite{chang2022distilhubert} and ALBERT \cite{lan2019albert} are chosen as backbones for visual, acoustic and textual modalities, respectively. The face images, raw audio array, and tokenized words pass through their respective backbones and 1D convolution layers (for capturing temporal information). Each embedding is then averaged to obtain representative modal features. To fuse these modalities, the averaged features are concatenated, and two linear layers are used to make the final prediction.

\begin{figure}[htb!]
\centering
\includegraphics[width=0.8\linewidth]{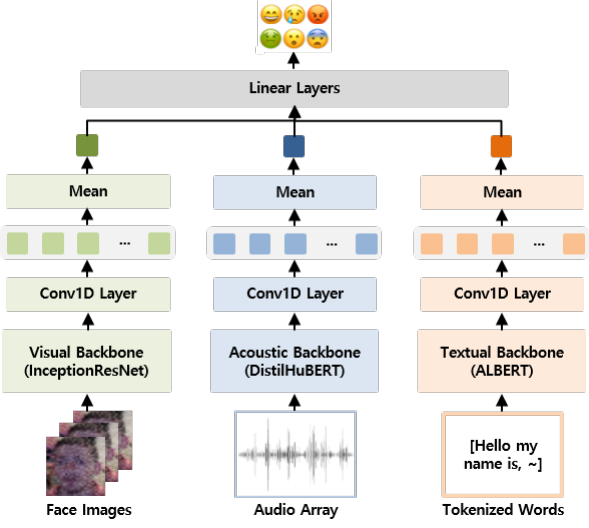}
   \caption{Overall design of the proposed MER model.}
\label{eemer}
\end{figure}

\subsection{Training and Results}
We trained the model on the CMU-MOSEI dataset \cite{zadeh2018multimodal}, one of the largest MER datasets available, which was restructured by \cite{dai2021multimodal}. The model is designed to take 5 face images, 10 seconds of audio and 100 tokens of words as inputs where they are either truncated or padded with 0 depending on their length. Since the dataset is multi-labeled, we optimized the model using binary cross-entropy loss and set thresholds for each emotion \cite{dai2021multimodal}. These thresholds were selected to maximize the F1 score in the validation set. Finally, the learning rates of the backbones were set to be 1/10 of the entire model. This way, the backbones can gradually adapt to the task of MER without totally forgetting the modal-specific knowledge. After training, the model achieved an F1 score of 48.3 and accuracy of 72.7 on the test set, outperforming the previous state-of-the-art results \cite{wei2022fv2es}.

\section{MER System}

\subsection{System Design}

\begin{figure}[htb!]
\centering
\includegraphics[width=0.9\linewidth]{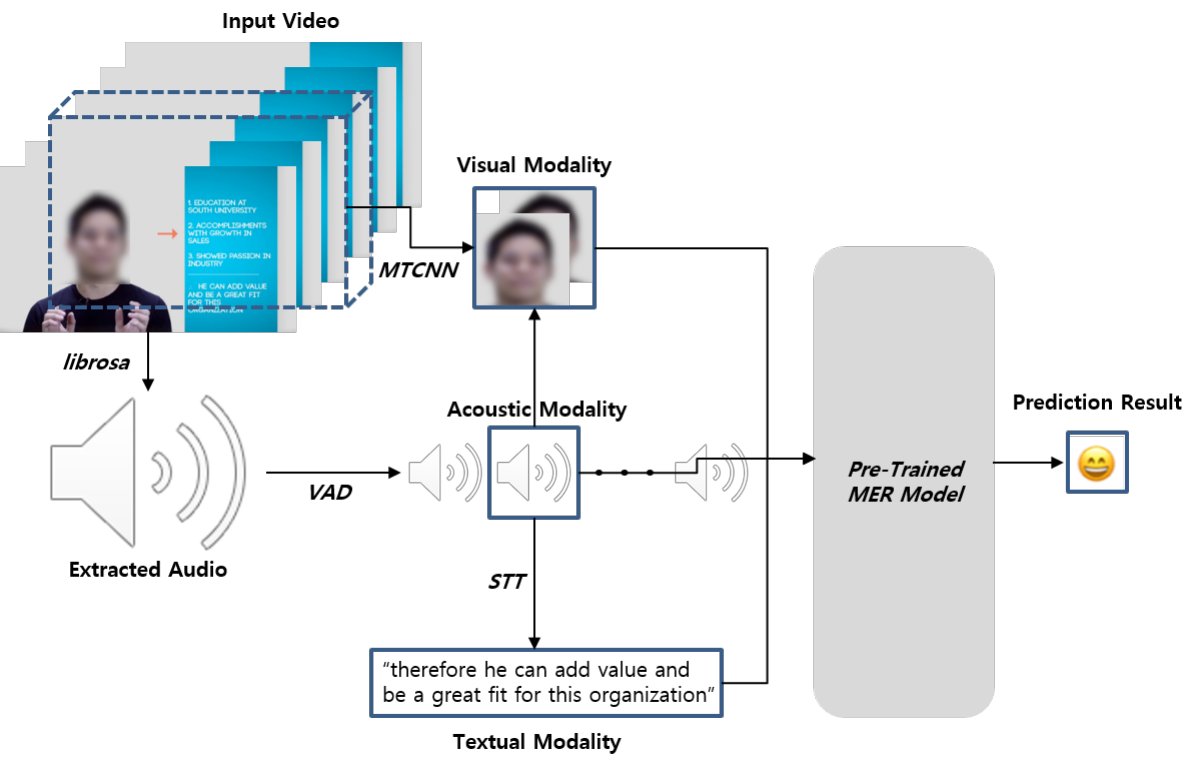}
   \caption{Overall flow of the proposed MER system.}
\label{system}
\end{figure}

Figure \ref{system} shows the overall flow of the proposed MER system. For each input video, the system utilizes the Python library librosa\footnote{https://librosa.org/doc/latest/index.html} to extract the audio information, and then applies an open source voice activity detection (VAD) model to obtain the timestamps of each utterance from the audio data. \textit{Silero VAD}\footnote{https://github.com/snakers4/silero-vad} was specifically chosen for this purpose in our system. According to the timestamps information, the utterance-level acoustic modality can be easily retrieved. For the utterance-level visual modality, 5 frames are sampled from all the frames and \textit{MTCNN}\footnotemark[1] is applied to crop face regions from them. As there is no direct way to get textual modality from the video, a speech-to-text model needs to be applied. We adopted \textit{Silero Models}\footnote{https://github.com/snakers4/silero-models} among various open source STT models. Now, with the information of the three extracted modalities, the pre-trained MER model can predict the speaker\textquotesingle{}s emotion at the utterance level. By averaging all the utterance-level emotion probabilities, the overall video-level emotions can also be recognized.

\subsection{System Interface}

\begin{figure}[htb!]
\centering
\includegraphics[width=0.9\linewidth]{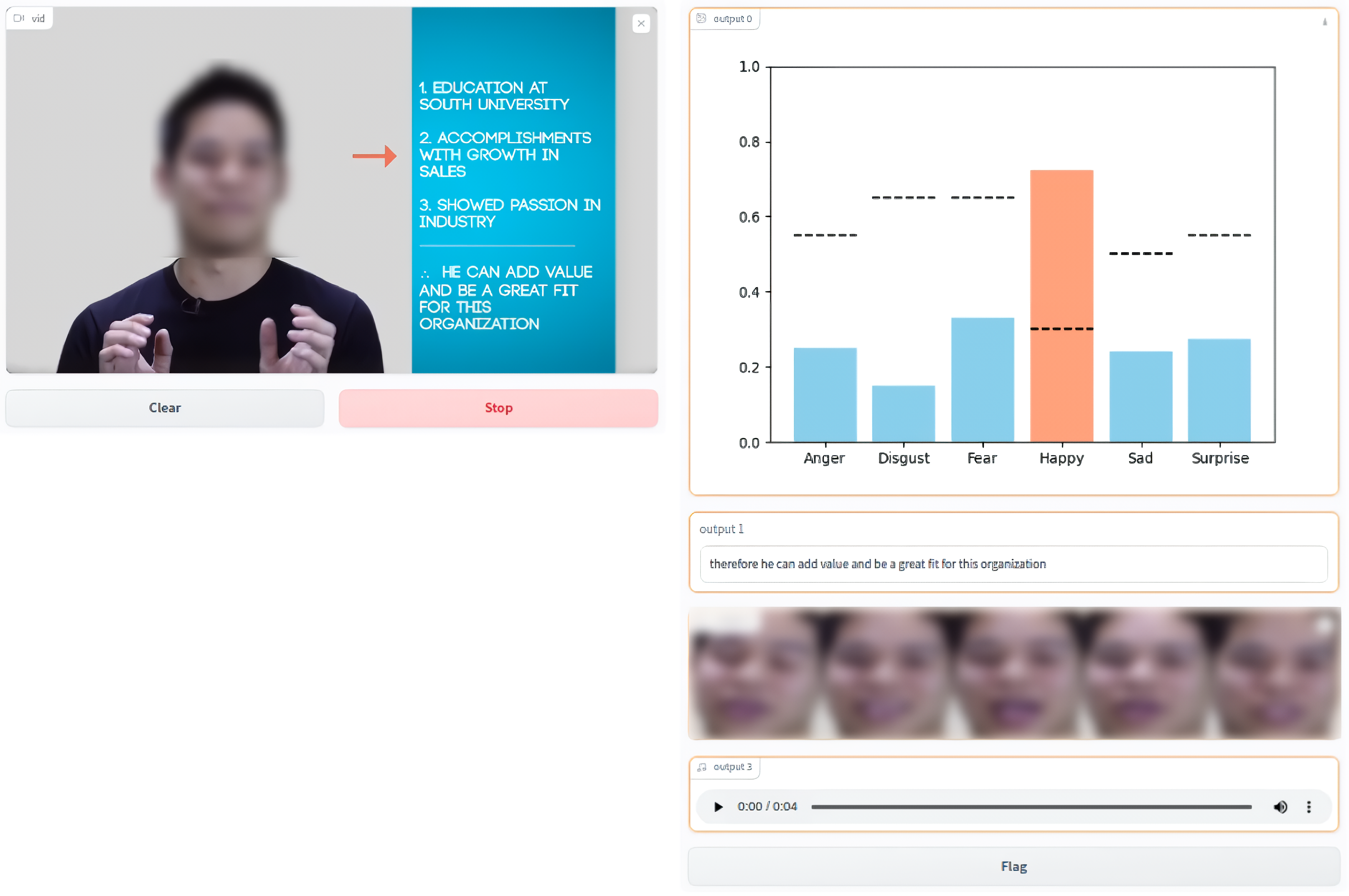}
   \caption{Overall structure of the MER system interface.}
\label{interface}
\end{figure}

The implemented system interface is shown in Figure \ref{interface}. In the left window, a user can upload a video. When clicking the \textit{Submit} button, the video is processed following the system description provided above. Firstly, the system processes the video and displays the utterance-level results sequentially in the right windows. The bar chart shows the prediction of the model and the dotted lines indicate the thresholds of each emotion. If the predicted value is above the threshold, the corresponding bar is highlighted with a different color to indicate that the emotion is expressed in the current utterance. Additionally, the three inputs that the model used for inference are shown below the bar chart. Once the processing is complete, the final result, which is the average of all the utterance-level predictions, is displayed. The user can then click the \textit{Clear} button to remove the current video and repeat the process with a new one.

\section{Conclusion}
In this paper, we presented a novel system for real world applications of MER in videos.The model at the core of the system is an end-to-end MER model based on large pre-trained models, which leverages the power of modal-specific knowledge to achieve state-of-the-art performance. This model is seamlessly integrated with other AI models such as VAD and STT to enable utterance-level emotion prediction for speakers in videos. Additionally, users can intuitively interact with the system through the provided system interface. Overall, we believe that our proposed system has the potential for widespread usability in diverse fields, such as healthcare, education, and entertainment.

\bibliographystyle{IEEEtran}
\bibliography{mybib}

\begin{thebibliography}{1}
\providecommand{\url}[1]{#1}
\csname url@samestyle\endcsname
\providecommand{\newblock}{\relax}
\providecommand{\bibinfo}[2]{#2}
\providecommand{\BIBentrySTDinterwordspacing}{\spaceskip=0pt\relax}
\providecommand{\BIBentryALTinterwordstretchfactor}{4}
\providecommand{\BIBentryALTinterwordspacing}{\spaceskip=\fontdimen2\font plus
\BIBentryALTinterwordstretchfactor\fontdimen3\font minus
  \fontdimen4\font\relax}
\providecommand{\BIBforeignlanguage}[2]{{%
\expandafter\ifx\csname l@#1\endcsname\relax
\typeout{** WARNING: IEEEtran.bst: No hyphenation pattern has been}%
\typeout{** loaded for the language `#1'. Using the pattern for}%
\typeout{** the default language instead.}%
\else
\language=\csname l@#1\endcsname
\fi
#2}}
\providecommand{\BIBdecl}{\relax}
\BIBdecl

\bibitem{wei2022fv2es}
Q.~Wei, X.~Huang, and Y.~Zhang, ``Fv2es: A fully end2end multimodal system for
  fast yet effective video emotion recognition inference,'' \emph{IEEE
  Transactions on Broadcasting}, 2022.

\bibitem{chang2022distilhubert}
H.-J. Chang, S.-w. Yang, and H.-y. Lee, ``Distilhubert: Speech representation
  learning by layer-wise distillation of hidden-unit bert,'' in \emph{ICASSP
  2022-2022 IEEE International Conference on Acoustics, Speech and Signal
  Processing (ICASSP)}.\hskip 1em plus 0.5em minus 0.4em\relax IEEE, 2022, pp.
  7087--7091.

\bibitem{lan2019albert}
Z.~Lan, M.~Chen, S.~Goodman, K.~Gimpel, P.~Sharma, and R.~Soricut, ``Albert: A
  lite bert for self-supervised learning of language representations,''
  \emph{arXiv preprint arXiv:1909.11942}, 2019.

\bibitem{zadeh2018multimodal}
A.~B. Zadeh, P.~P. Liang, S.~Poria, E.~Cambria, and L.-P. Morency, ``Multimodal
  language analysis in the wild: Cmu-mosei dataset and interpretable dynamic
  fusion graph,'' in \emph{Proceedings of the 56th Annual Meeting of the
  Association for Computational Linguistics (Volume 1: Long Papers)}, 2018, pp.
  2236--2246.

\bibitem{dai2021multimodal}
W.~Dai, S.~Cahyawijaya, Z.~Liu, and P.~Fung, ``Multimodal end-to-end sparse
  model for emotion recognition,'' in \emph{Proceedings of the 2021 Conference
  of the North American Chapter of the Association for Computational
  Linguistics: Human Language Technologies}, 2021, pp. 5305--5316.

\end{thebibliography}

\end{document}